\documentclass{amsart}

\usepackage{amsmath}
\usepackage{latexsym}
\usepackage{amsfonts}
\usepackage{amssymb}
\usepackage{color}
\usepackage{bbm,dsfont}
\usepackage{graphicx}
\usepackage{textcomp}
\usepackage{enumerate}
\usepackage{MnSymbol}
\usepackage{bbding}
\usepackage{multirow}
\usepackage{array}
\usepackage{makecell}
\usepackage[table]{xcolor}

\theoremstyle{plain}

\theoremstyle{definition}

\theoremstyle{remark}

\newcommand{\boldv}{{\vrule width 1.2pt}} 
\newcommand{\boldh}{\Xhline{1.2pt}} 

\newcommand{\half}{\tfrac{1}{2}} 

\newcommand{\kb}[2]{|#1\rangle\langle#2|} 
\newcommand{\tr}[1]{\mathrm{tr}\left[#1\right]} 
\newcommand{\id}{\mathbbm{1}} 

\newcommand{\Mo}{\mathsf{M}}
\newcommand{\No}{\mathsf{N}}
\newcommand{\Po}{\mathsf{P}}
\newcommand{\Xo}{\mathsf{X}}
\newcommand{\Yo}{\mathsf{Y}}
\newcommand{\Zo}{\mathsf{Z}}

\newcommand{\nrank}{\mathrm{rank}_+} 
\newcommand{\psdrank}{\mathrm{rank}_{psd}} 

\begin{document}

\title{Can a qudit carry more information than a dit?}


\author{Teiko Heinosaari}

\author{Mark Hillery}


\maketitle

\begin{abstract}
Conventional wisdom suggests that within a fixed preparation-measurement setup, a qubit system offers no advantage over a bit.
This indeed holds true when considering the standard communication and the famous Holevo bound then formalizes the statement that one qubit can encode at most one bit of information.
However, there exist subtle differences between these two physical systems that, when properly exploited, can be converted into practical applications.
We begin by discussing the similarities between qudits and dits as information carriers.
Then we recall a general framework for communication tasks and review some differences that qudits and dits have.
In the end, we present a simple communication application that utilizes the quantum character of the qubit.
\end{abstract}

\maketitle

\section{Introduction}\label{sec:intro}

Quantum technology is a multi-disciplinary and rapidly developing field that has impacts on various technologies, most notably in information processing.
The advance of quantum information and computation ultimately rests on the identification of quantum resources and then finding ways to utilize them ingeniously and efficiently.
In that way, one may be able to find ways to circumvent some limitations of classical information processing.
Successful applications often use quantum entanglement as their primary quantum resource \cite{spiller2005introduction}.
In quantum communication, there are several protocols that benefit from entanglement. 
A simple yet fascinating example is superdense coding, which is a way to transmit information coded in two bits by sending only one bit, assuming that the communication parties share an entangled pair of qubits \cite{bennett1992communication}. In quantum computing, there is evidence that entanglement is crucial for quantum speedup, although the role of entanglement in computational speedup is not completely clear yet \cite{HoHoHoHo09}.
Apart from applications, an important general theorem is the celebrated Holevo bound \cite{Holevo73}, which expresses the fact that from one qubit one can retrieve at most one bit of information.

The previous points suggest that quantum applications must use composite systems in one way or another in order to make a difference when compared to the respective implementation with systems obeying classical physics.
Classical and quantum systems are evidently very different kind, but is it really so that a single qubit is no better than a bit? Or more generally and precisely, is a single quantum system of a fixed dimension more useful in some information processing task than a classical system with the same dimension? If so, what is the advantage that a single quantum system can give?

It should be perhaps emphasized that `being better' and `being dissimilar' are two different questions. 
Fundamental phenomena of quantum physics that cannot be understood in classical terms have been investigated from various viewpoints \cite{jennings2016no} and the non-classicality of a qudit is a fact.
Further, there have been activities in new axiomatizations of quantum theory and basically all of them provide some viewpoint to `quantumness' of a qudit by giving critical axioms that quantum theory does satisfy but classical theory does not \cite{chiribella2016quantum}.
In the following our interest is in simple information transmission tasks that have clear operational motivation and can be equally formulated both in classical and in quantum settings.
This kind of operational framework can be thought as a test where a referee is testing abilities of two parties (Alice and Bob) to send information in some specified way. Alice and Bob can use either a dit or a qudit in their planned act, and the performances in these two cases are compared.

The purpose of this contribution is to give an easily accessible review of the main arguments and results of how a single qubit is in some respects not different from a bit, while there are some information processing tasks that reveal their differences. Some naturally arising open questions are also discussed. We start by Section \ref{sec:ditqudit} with a brief summary of classical and quantum descriptions of states and measurements.
In Section \ref{sec:same} we go through three basic tasks where quantum systems give no advantage over classical systems, although one could first believe the opposite.
In Section \ref{sec:different} we inspect some other tasks where the difference between classical and quantum becomes visible.
By exploiting one of the quantum features of qubits, we introduce a novel communication application in Section \ref{sec:random}.
Finally, in Section \ref{sec:conclusion} we present some concluding remarks.

\section{Dit and Qudit}\label{sec:ditqudit}

In this section we recall the basic mathematical formalism of finite dimensional quantum systems.

\subsection{Classical and quantum state spaces}\label{sec:states}

A dit is a classical system with exactly $d$ distinguishable states.
For instance, a bit system ($d=2$) has two pure states and we can denote them 
\begin{align}
\left(\begin{array}{c}1 \\ 0\end{array}\right) \, , \quad \, \left(\begin{array}{c}0 \\ 1\end{array}\right) \, .
\end{align}
These states could e.g. correspond to the red and green light signals of a pedestrian traffic light.  
For that purpose it would be enough to just use 0 and 1, but the vector notation makes it possible to have mixed states.
A general bit state is a mixture of these two states and it has the form
\begin{align}\label{eq:bit-mixed-state}
t \cdot \left(\begin{array}{c}1 \\ 0\end{array}\right) + (1-t) \cdot \left(\begin{array}{c}0 \\ 1\end{array}\right) = \left(\begin{array}{c}t \\ 1-t\end{array}\right) 
\end{align}
for some $t\in [0,1]$. 
For classical state spaces, mixtures typically describe uncertainty of an observer to know the actual state.
Summarizing,  the state space of a dit is the set of all $d$-component probability vectors.

A $d$-dimensional quantum system is called a qudit.
The dimension refers to the maximal number of simultaneously distinguishable states;
a qudit system has infinitely many pure states, but only $d$ of those can be simultaneously distinguishable. 
The qudit state space consists of positive $d \times d$ matrices with unit trace (i.e. all eigenvalues are non-negative and sum to $1$) and these are also referred as density matrices.
The extreme elements of the convex set of all qudit states are called pure states and they correspond to 1-dimensional projections.
For instance, a general qubit state is a $2\times 2$ matrix
\begin{align}
\varrho = \left(\begin{array}{cc} \varrho_{00} & \varrho_{01} \\ \varrho_{10} & \varrho_{11} \end{array}\right)
\end{align}
that has nonnegative eigenvalues and satisfies $\tr{\varrho}=1$. 

When we write qudit states as matrices, we have implicitly fixed an orthonormal basis of the underlying complex Hilbert space.
Naturally, we can choose another basis and then a matrix $\varrho$ becomes $U^* \varrho U$, where $U$ is the unitary operator corresponding to the change of the bases.
The important point is that by fixing an orthonormal basis one can see the dit states as specific qudit states.
Namely, a probability vector determines a diagonal density matrix uniquely.
For instance, the classical bit states written in \eqref{eq:bit-mixed-state}
can be seen as diagonal qubit states as
\begin{align}
\left(\begin{array}{cc} t & 0 \\ 0 & 1-t \end{array}\right) \, .
\end{align}
The interpretation is that a qudit state space has a copy of a classical state space inside of it, corresponding to the restriction to use the pure states in some fixed orthonormal basis and their mixtures.

The feature that the classical state space is missing is coherence, or in other words, possibility to form superpositions.
Coherence is a critical quantum resource in decision problems, such as the Deutsch-Jozsa algorithm \cite{hillery2016coherence}.
In the convex structure of quantum state space the possibility to form superpositions is reflected as non-unique convex decompositions of mixed states \cite{HuJoWo93}. 
This is in contrast to mixed classical states, which have unique convex decompositions into pure states.

\subsection{Quantum measurements}

A standard quantum measurement on a qudit is determined by an orthonormal basis.
Namely, an orthonormal basis $\{\varphi_0,\ldots,\varphi_{d-1}\}$ defines an operator-valued function $\Po$ via the relation
\begin{align}\label{eq:sharp}
\Po(j)=\kb{\varphi_j}{\varphi_j} \, .
\end{align}
This function is a mathematical description of the measurement and the interpretation connecting mathematics to physics is that the probability of obtaining a measurement outcome $j$ when the initial state is $\varrho$ is given by the Born formula, 
\begin{align}
\textrm{Prob}(j)=\tr{\varrho \Po(j)}  \, .
\end{align}
The operators $\Po(j)$ are one-dimensional projections and hence this measurement is also referred as a projective measurement.
Performing a projective measurement `reads one dit' from a qudit.

Already a qubit system has infinitely many different orthonormal bases and each of them corresponds to a different way of reading one bit from the qubit.
The three qubit measurements often used in quantum computation are called $\Xo$-, $\Yo$- and $\Zo$-measurements and we denote their outcomes as $\pm$.
It is common to write the measurements in the $\Zo$-eigenbasis, so that 
\begin{align}
\Zo(+)=  \left(\begin{array}{cc} 1 & 0 \\  0 & 0 \end{array}\right) \, , \quad \Zo(-)=  \left(\begin{array}{cc} 0 & 0 \\  0 & 1 \end{array}\right) \, ,
\end{align}
while
\begin{align}
\Xo(\pm)= \frac{1}{2} \left(\begin{array}{cc} 1 & \pm 1 \\ \pm 1 & 1 \end{array}\right) \, , \qquad
\Yo(\pm)= \frac{1}{2} \left(\begin{array}{cc} 1 & \mp i \\ \pm i & 1 \end{array}\right) \, .
\end{align}
The fundamental feature of these quantum measurements is that they cannot be measured simultaneously.
Their joint measurement becomes possible only by accepting a certain amount of unsharpness \cite{Busch86}.
This quantum incompatibility is the fundamental feature of collections quantum measurements \cite{HeMiZi16}, similarly as coherence is the fundamental feature of monopartite quantum states and entanglement of multipartite quantum states. 
All these crucial quantum resources - coherence, entanglement, and incompatibility - are vulnerable to noise, which can result in the dissipation of quantum characteristics of a system.

An important point for our later discussion is that a measurement on a qudit can have more than $d$ outcomes.
The way to implement such a measurement is that we couple the qudit to another ancillary qudit in a fixed state, operate them with a global unitary transformation and finally make a projective measurement on the composite $d^2$-dimensional system. 
As a result, we have realized a $d^2$-outcome measurement that can be interpreted as a measurement on the original qudit.
It is not anymore a projective measurement, but described as a positive operator valued measure (POVM): to each outcome $j$ is assigned a positive operator $\Mo(j)$ and these operators sum to the identity operator, i.e., $\sum_j \Mo(j) = \id$. 
The operators are determined from the equation
\begin{align}\label{eq:dilation}
\tr{\varrho \Mo(j)} = \tr{U(\varrho \otimes \varrho_0)U^* \Po(j)} \, , 
\end{align}
required to hold for all states $\varrho$.
Even if $\Po$ in \eqref{eq:dilation} is a projective measurement, the POVM $\Mo$ can have noncommuting elements.
One can also go backward: any measurement with $d^2$ or less outcomes has a realization in the form of \eqref{eq:dilation}.
This is known as the Naimark's theorem.

Once we have accepted that quantum measurements are mathematically described as POVMs, there is no upper limit for the number of measurement outcomes. 
It is possible even to have qubit measurements with a continuous number of measurement outcomes \cite{MLQT12}.
However, in finite dimension a quantum measurement with a continuous set of outcomes can be written as a continuous random choice of measurements with a finite number of outcomes \cite{ChDaSc07}. Further, $d^2$ is the critical bound in the sense that every measurement with finite number of outcomes can be obtained from some collection of measurements with $d^2$ or fewer outcomes by mixing them and relabeling the outcomes \cite{HaHePe12}.
An additional aspect arising from the procedures of mixing and relabeling is that we do not necessarily need any ancillary system in order to implement measurements with more than $d$ outcomes \cite{guerini2017operational}. 
These kind of measurements formed by mixing and relabeling can be useful, even if not all measurement have that kind of implementation \cite{FiHeLe18}.

To provide a concrete example of mixing and relabeling procedures, suppose we are measuring a qubit system and at each measurement round we choose randomly either $\Xo$ or $\Yo$ measurement. 
We record the obtained outcome and also keep track of the chosen measurements.
In this way, there are four possible outcomes: $\pm_\Xo$, $\pm_\Yo$.
If the probabilities of choosing $\Xo$ and $\Yo$ are equal, then the total four outcome measurement $\Mo$ is given as
\begin{align}
\Mo(j_\Xo) = \half \Xo(j) \, , \quad \Mo(j_\Yo) = \half \Yo(j)
\end{align}
for $j=\pm$.
We can further relabel the outcomes so that $0_\Xo$ becomes $0$, $0_\Yo$ becomes $1$ and the remaining outcomes $1_\Xo$ and $1_\Yo$ are merged into a single outcome that we label $?$.  The three outcome measurement formed in this way is then
\begin{align}
\No(0) =\half \Xo(+) \, , \quad \No(1) =\half \Yo(+) \, , \quad \No(?) = \half \Xo(-) + \half \Yo(-)  \, .
\end{align}
The measurement $\No$ has a feature that neither $\Xo$ nor $\Yo$ separately has. 
Suppose that we are given an unknown state that is either $\varrho_{x+}$, $\varrho_{x-}$, $\varrho_{y+}$ or $\varrho_{y-}$, where these are the eigenstates of the corresponding operators. 
If we get the outcome 0 in $\No$-measurement, we can conclude that the input state was not $\varrho_{x-}$. 
Similarly, if we get the outcome 1, we know that the input state was not $\varrho_{y-}$.
In the case of getting the outcome ? we cannot make any conclusion.

\section{In what sense dit and qudit are similar?}\label{sec:same}

In the following, by a message we mean a symbol (or letter) chosen from a fixed alphabet, and an alphabet is a finite set of symbols.
Let us consider a task where Alice tries to transmit a message to Bob by sending a single physical system, either dit or qudit. 
The physical system is the carrier of information and messages must be encoded into the states of the system in some way.
Both dit and qudit have exactly $d$ perfectly distinguishable states, which means that a message can be encoded and decoded in a reversible manner if there are $d$ elements in the alphabet, at most. 
Since a qudit has infinitely many pure states compared to $d$ pure states of a dit, one may wonder whether this abundance of pure states can be used in favour of a qudit in some related communication scenario. 
To investigate this possibility, let us assume that the alphabet has $N$ letters, with $N>d$. 
This causes an issue for communication as we do not have enough perfectly distinguishable states, neither with dit nor with qudit.
Hence, the success rate of transmission of information must be quantified in some reasonable way in order to compare the two physical systems.
We can repeat the trial many times and quantify the overall result by determining the success probability of being able to decode the message correctly.
Since a qudit system has more possible pure states as a classical dit system, it might be that there is a way to benefit from this fact.

To define certain communication tasks more precisely, let us assume that the referee randomly draws one letter from the alphabet. 
The referee announces the letter to Alice, who then has to transmit the message to Bob. 
The essential point is that Alice can send only one physical system to Bob and our interest is to compare the two situations of communication medium being dit and qudit.
After receiving the physical system, Bob performs a measurement and tells his guess to the referee.
There are two common variants of this communication task, depending whether Alice and Bob aim to minimize the error, or if error is not tolerated at all but then it is accepted that there are rounds with inconclusive outcome. The latter option means that Bob tells 'I don't know the message' to the referee.
We consider these two variants separately.

\subsection{Minimum error communication}\label{sec:med}

As a starting example, let us consider the task of transmitting $x\in\{a,b,c\}$ with a single bit. The input $x$ is drawn with the uniform probability.
Before the communication test begins, Alice and Bob can agree on their encoding-decoding startegy.
For instance, Alice and Bob can decide that Alice encodes $a$ to $0$ and $b$ to $1$. With $c$ there are no more pure states left and they decided that Alice encodes also $c$ to $0$. 
It is then clear that Bob will make errors when trying to decode the message. 
Bob obviously decodes $1$ to $b$, but $0$ can mean either $a$ or $c$. 
He can choose to decode $0$ half of the time to $a$ and half of the time to $c$, the selection being made by tossing a coin at each round.
With this strategy the transmission of information succeeds with the probability $2/3$.
In this simple case it is possible to make a table with all other strategies and one can confirm that any strategy leads to the success probability $2/3$, at most.

More generally, we can consider a communication test with an alphabet of size $N$ and a dit as a communication medium.
An analogous strategy as described previously leads to the success probability $d/N$.
One can also try different kind of encoding and decoding scenarios, but they never give better success probability than $d/N$.
The fact that this is an absolute upper bound is a consequence of the following consideration.

Let us inspect the same task but now with the transmission of a single qudit.
For each possible message $x$ is assigned a qudit state $\varrho_x$ and for the decoding there is a $N$-outcome measurement $\Mo$.
The success probability is
\begin{align}\label{eq:bdt-1}
p_{\textrm{ME}}=\frac{1}{N} \sum_{x=1}^N \tr{\varrho_x \Mo(x)} \, .
\end{align}
For a fixed measurement $\Mo$, the best encoding state is such that $\tr{\varrho_x \Mo(x)} = m_x$, where $m_x$ is the greatest eigenvalue of $\Mo(x)$. 
This choice of states gives $p_{\textrm{ME}} =\frac{1}{N} \sum_x m_x$.
Using the normalization condition $\sum_x \Mo(x) = \id$ we observe that
\begin{align}
\sum_x m_x \leq \sum_x \tr{\Mo(x)} =  \tr{\sum_x\Mo(x)} = \tr{\id}=d \, .
\end{align}
Therefore, we conclude that for all choices of encoding states and decoding measurement there is the upper bound
\begin{align}\label{eq:me-bound}
p_{\textrm{ME}} \leq \frac{d}{N} \, ,
\end{align}
valid for all $N \geq d$.
This bound is known as the Basic Decoding Theorem \cite{QPSI10}.
Since the upper bound is already achieved by using a dit, we come to the conclusion that a single qudit is no better than a dit in transmitting information in the described minimum error valuation.

\subsection{Unambiguous communication}\label{sec:uad}

Let us then move to the other variant of communication, related to unambiguous discrimination of states. 
As before, the referee announces a symbol $x\in \{a,b,\ldots\}$ to Alice and 
Alice can send one dit or qudit to Bob, depending on the considered scenario. 
In the unambiguous communication task, Bob can say that he doesn't know the symbol, but when he announces the symbol he is not allowed to make a mistake. 
The success probability is still defined as the sum
\begin{align}\label{eq:uad-1}
p_{\textrm{UA}}=\frac{1}{N} \sum_{x} \ \tr{\varrho_x \Mo(x)} \, ,
\end{align}
but now with the constrains 
\begin{equation}\label{eq:uad-con-1}
\begin{split}
& \tr{\varrho_x \Mo(y)}=0 \quad \textrm{for $x\neq y$} \\
& \sum_x \Mo(x) \leq \id 
\end{split}
\end{equation}
as Bob should not make any mistakes and there is one extra outcome for the inconclusive answer $?$.

The interesting tasks are again those where the number of symbols $N$ is greater than the operational dimension $d$ so that there are not enough perfectly distinguishable states for the encoding.
We further note that the unambiguous transmission task of $N$ symbols is at least as hard as the corresponding minimum error transmission task in the sense that $p_{\textrm{UA}} \leq p_{\textrm{ME}}$, assuming that we use the same states and optimize the measurements for unambiguous and minimum error tasks, respectively.
This inequality simply follows from the fact that we can translate any unambiguous measurement $\Mo$ to a minimum error measurement $\Mo'$ by randomly choosing some label whenever the extra outcome $?$ is obtained. 
Mathematically, $\Mo'(x)=\Mo(x) + \tfrac{1}{N} \Mo(?)$ for all $x$ and this implies that $\tr{\varrho_x \Mo'(x)} \geq \tr{\varrho_x \Mo(x)}$.
Therefore, the upper bound \eqref{eq:me-bound} is also an upper bound for the success probability in unambiguous communication. 
It was noticed in \cite{Chefles98} that pure quantum states can satisfy the unambiguous discrimination task with $\tr{\varrho_x \Mo(x)}>0$ only if the corresponding vectors in Hilbert space are linearly independent. In the following we develop this observation to get a tighter upper bound for $p_{\textrm{UA}}$ than \eqref{eq:me-bound}.

Let us suppose that Alice transmits qudits to Bob and she is using $N$ pure states in the encoding, with $N>d$.
The pure states correspond to unit vectors $\psi_1,\ldots,\psi_N$ and \eqref{eq:uad-con-1} gives
\begin{align}\label{eq:uad-vectors}
\Mo(y)\psi_x=0 \quad \textrm{for $x\neq y$} \, .
\end{align}
The set of all $N$ vectors cannot be linearly independent, and any nontrivial linear constraint implies that the corresponding indices do not contribute to the success probability.
Namely, suppose e.g. that $\psi_1=c_2 \psi_2 + c_3 \psi_3$  for some $c_2 \neq 0 \neq c_3$.
Since from \eqref{eq:uad-vectors} we get $\Mo(1)\psi_2=0$ and $\Mo(1)\psi_3=0$, it follows that $\Mo(1)\psi_1=0$.
Further, the linear equation can be rewritten so that $\psi_2$ and $\psi_3$ are linear combinations of the other two vectors, therefore we conclude in a similar way that $\Mo(2)\psi_2=0$ and $\Mo(3)\psi_3=0$.
The same reasoning applies to any linear constraint with nonzero coefficients.
Therefore, to maximize the success probability, we need to use as many linearly independent vectors as possible, which is $d$. 
The remaining $N-d$ vectors are necessarily linear combinations of those $d$ vectors. 
As a linear constraint implies that all involved vectors do not contribute to the success probability, we want the $N-d$ vectors to be linear combinations of as few vectors as possible. 
Hence, this leads to choosing them to be the same vector as one of the $d$ vectors.
This means that $N-d+1$ symbols are encoded to the same pure state and that has then to be assigned to the inconclusive outcome ?.
In the sum \eqref{eq:uad-1} there are at most $d-1$ terms that are nonzero and each of those is at most $1$, thus we obtain the upper bound
\begin{equation}\label{eq:ua-bound}
p_{\textrm{UA}} \leq \frac{d-1}{N} \, ,
\end{equation}
valid for all $N>d$.
We note that using mixed states instead of pure states cannot lead to a higher success probability as the zero probability condition for $x\neq y$ becomes just harder to satisfy. 

Similarly, as in the case of minimum error communication, the upper bound \eqref{eq:ua-bound} can be reached already by using a dit as a communication medium.
Namely, we choose $d$ perfectly distinguishable states $\varrho_0,\ldots,\varrho_{d-1}$ (corresponding to orthogonal unit vectors) and set the other $N-d$ states to be the same as $\varrho_{d-1}$, i.e., $\varrho_{d-1}=\varrho_d = \ldots=\varrho_{N-1}$.
In the decoding measurement, an outcome $x$ gives a conclusive result for the first $d-1$ symbols while all other outcomes are assigned to the inconclusive outcome $?$.
This gives the maximal value $p_{\textrm{UA}} = (d-1)/N$.

\subsection{Communication of partial ignorance}

Minimum error and unambiguous communication tasks are specific types of communication tasks.
An interesting concrete example of another kind of task is communication of partial ignorance.
In the general formulation of this task, there are $N>d$ possible symbols and at each round the referee randomly chooses one of them.
The referee does not tell his choice to Alice but he randomly chooses another symbol and tells this wrong symbol to Alice. 
In this way, Alice does not know the correct symbol but she is not anymore totally ignorant as she was in the beginning - that's why it is referred as partial ignorance \cite{HeKe19}.
The collaborative aim of Alice and Bob is the same as before: Alice transmits information to Bob and based on that information Bob tries to guess the correct symbol (i.e. the first choice of the referee). 
We count the success probability of Bob guessing the correct symbol. It is clear that Alice and Bob cannot succeed always since even Alice does not know the correct symbol.
Therefore, the best what they can aim for is that the communication step does not decrease Bob's guessing probability when compared to the situation where Alice makes a guess alone without any communication.
Alice knows one of the wrong symbols and hence she guesses the correct symbol with the probability $1/(N-1)$. 
This is taken to be the value that indicates faultless communication of partial ignorance and we conclude that the success probability $p_{\textrm{PI}}$ in the communication of partial ignorance has the upper bound
\begin{equation}\label{eq:pi-bound}
p_{\textrm{PI}} \leq \frac{1}{N-1} 
\end{equation}
since clearly no communication can help Alice and Bob to make a better guess than Alice alone.

In searching for optimal communication strategies, let's first assume that Alice can send a bit to Bob. 
Alice and Bob can agree that if the wrong symbol told by the referee to Alice is $a$, then Alice sends $0$ and in the other cases Alice sends $1$ to Bob. If Bob receives $1$, he announces $a$ to the referee. If he receives $0$, he makes a random choice between the $N-1$ labels different from $a$. With this strategy, Alice and Bob make full use of the information available to them in the sense that Bob never announces the symbol that the referee told to be wrong.
In fact, by a straightforward calculation one can verify that Alice and Bob reach the success probability  $1/(N-1)$ with their strategy.

A notable difference of this task of transmitting partial ignorance to the earlier two discrimination tasks is that the bound \eqref{eq:pi-bound} does not depend on the dimension $d$, and the earlier classical strategy using one bit gives the maximal success probability for any $N$.
It may hence seem like this task would be most far from revealing some advantage when a quantum system is being used.
Surprisingly, in the next section we explain that a small modification in the transmission task of partial ignorance makes a critical difference.
 
\section{In what sense bit and qubit are different?}\label{sec:different}

\subsection{Communication of uniform partial ignorance}

Let's still consider the setup of the communication of partial ignorance.
As we saw earlier, with the classical strategy Alice and Bob make full use of the information available to them in the sense that Bob never announces the symbol that the referee told to be wrong.
However, Bob makes the guess $a$ more often than the other options, so it seems as Alice's uniformly distributed ignorance about the correct symbol has not been transmitted to Bob in a faithful way. 
In fact, as pointed out in \cite{HeKe19}, one modification in the task makes the situation different and reveals a separation between bit and qubit.
Suppose that the test is run many times and the referee is observing the strategy that Alice and Bob are using. 
The referee can try to minimize the guessing probability of Bob by choosing non-uniformly the correct symbol and the announced wrong symbol. For instance, if in the previous example the referee chooses $b$ as the correct symbol and $c$ as the announced wrong symbol, then Bob never guesses $b$. Hence, Alice and Bob would always fail. 

With a qubit Alice can communicate the announced wrong answer in a different way, where each wrong possibility is assigned to a distinct state. 
These states must be chosen in a way that when Alice encodes $a$, Bob will never get $a$ with his measurement, and similarly for the other symbols.
The corresponding property of quantum states is called antidistinguishability or antidiscrimination. 
A qubit system has collections of antidistinguishable states for any $N$ \cite{HeKe18}.
As we already noted, in terms of the success probability $p_{\textrm{PI}}$, these quantum states do not give any advantage with the classical strategy that we explained earlier. However, if they are chosen in a particular way, they make the strategy immune to the referee's harmful actions. 
Alice and Bob need to choose a preparation-measurement setup such that if Alice sends a state $\varrho_x$, then Bob's measurement gives the outcome $x$ with zero probability and other indices with the equal probability $1/(N-1)$.
The underlying reason why this setup can circumvent the classical limitation (i.e. sensitivity to the referee's evil action) is that Alice cannot control what outcome Bob gets, apart from the fact that it is not $x$. 
In fact, Alice does not even know which wrong option Bob will exclude after their communication - she can just be sure that Bob excludes one of the wrong options and makes a uniform guess among the other options.
Using a qubit, Alice and Bob have a preparation-measurement setup with the required property for $N=3$ and $N=4$, but not for any higher $N$ \cite{HeKe19}.
The qubit states in these setups correspond to uniformly distributed vectors in the Bloch sphere, hence to a triangle and a tetrahedron, respectively.
The measurements are defined via the condition that the states are in the kernel of the corresponding operators, i.e., if $\varrho_x$ is the input state that encodes that $x$ is a wrong answer, then the corresponding measurement operator is $\Mo(x)=c(\id - \varrho_x)$, where $c$ is a normalization constant chosen such that $\sum_x \Mo(x) = \id$.
The specific geometric constellation of the states implies that $\tr{\varrho_x \Mo(y)} = 1/(N-1)$.

\subsection{Full variety of communication tasks}

We have discussed three different communication tasks. They are still special forms of communication, and we might be come up with other such tasks.  
A method to talk about all possible communication tasks is to formulate them as matrices.
This kind of matrix is nothing else than a table of desired probabilities for preparation and measurement outcome pairs. 
If we list preparations in the rows and measurement outcomes in the columns, then the matrix is row-stochastic.
For instance, the perfect communication of three symbols corresponds to the identity matrix
\begin{equation}\label{eq:I3}
I_{3} = \left[\begin{array}{ccc} 1 & 0 & 0  \\0 & 1 & 0  \\ 0 & 0 & 1 \end{array}\right] \, ,
\end{equation}
while the previously explained uniform antidistinguishability of three symbols corresponds to the matrix
\begin{equation}\label{eq:A3}
A_{3} = \frac{1}{2} \left[\begin{array}{ccc} 0 & 1 & 1  \\1 & 0 & 1  \\ 1 & 1 & 0 \end{array}\right] \, .
\end{equation}
More generally, uniform antidistinguishability of $n$ symbols corresponds to the matrix
\begin{equation}\label{eq:D-n-1}
A_{n} = \frac{1}{n-1} \left[\begin{array}{ccccc} 0 & 1 & 1 & \cdots & 1 \\1 & 0 & 1 & \cdots & 1 \\ 1 & 1  & 0 & & 1 \\ \vdots & &   & \ddots  \\ 1 & \cdots & \cdots & 1 & 0 \end{array}\right] \, .
\end{equation}
This matrix describes a communication scenario where Alice communicates the wrong answer among $n$ possible symbols to Bob so that Bob get one of the other outcomes with uniform probability.
Other concrete examples can be found from \cite{heinosaari2020communication}.

We can accept the viewpoint that any row-stochastic matrix describes some communication task, whether or not it is known to be useful or not.
The question is then if a given communication matrix has an implementation with a qudit or with a dit as a communication medium.
A communication matrix $C$ with the size $m\times n$ has an implementation with a qudit system if there $m$ qudit states $\varrho_1,\ldots,\varrho_m$ and $n$-outcome qudit measurement $\Mo$ such that 
\begin{align}\label{eq:Cxy}
C_{xy}=\tr{\varrho_x \Mo(y) } 
\end{align}
for all $x=1,\ldots,m$ and $y=1,\ldots,n$.
An implementation with a dit system is equivalent to requiring that the states and effects in \eqref{eq:Cxy} are diagonal with respect to some fixed basis. 

Any communication matrix can be implemented with a high enough dimensional classical (hence also quantum) setting, therefore it is reasonable to ask for the minimal classical and quantum dimensions needed in the implementation.
These turn out to be equivalent to calculating the mathematical quantities called nonnegative rank (denoted $\nrank$) and positive semidefinite rank (denoted $\psdrank$) of the matrix, respectively.
In mathematical literature, they are defined slightly differently, but with some manipulation one can show that on row-stochastic matrices they agree with the classical and quantum minimal dimensions \cite{LeWeWo17}, respectively.
These quantities are thus giving precise mathematical formulation for the task at hand but, unfortunately, they are difficult to compute \cite{vavasis2010complexity,shitov2017complexity}.
Or, if we think positively, the computational difficulty makes the question interesting.
By the definition of these functions, they satisfy $\psdrank(C) \leq \nrank(C)$. 
This is nothing else than a mathematical expression for the fact that the minimal quantum dimension cannot be larger than the minimal classical dimension, which is obvious as we can interpret a $d$-dimensional classical system as a specific kind of $d$-dimensional quantum system.
The difference between dit and qudit is reveal in communication matrices $C$ having $d=\psdrank(C) < \nrank(C)$. 
For instance, $\psdrank(A_3)=2$ while $\nrank(A_3)=3$, and this means that the previously discussed uniform partial ignorance cannot be transmitted with a bit system although it is possible with a qubit.

If the quantum dimension of a given communication matrix can be smaller than the classical dimension, then how big can the separation be?
It has been recently shown in \cite{heinosaari2024simple} that already with a qubit system we can implement communication matrices that would need arbitrarily large classical dimension, i.e., there is a sequence of communication matrices $\{ C_k \}_{k=1}^{\infty}$ such that $\psdrank(C_k)=2$ for all $k$ and $\nrank(C_k) \geq k$.
In other words, there is a collection of communication tasks that can be implemented with a qubit system but which cannot be implemented by any classical system of fixed size.
The matrices used in the proof \cite{heinosaari2024simple} of this statement are related to antidistinguishability, but they are not uniform antidistinguishability matrices as qubit can implement only $A_3$ and $A_4$.
An outstanding open question is to understand the practical aspect of this kind of unbounded advantage.
In the next section we explain the advantage that $A_3$ can give.

\section{Application: random sequences with non-overlaps}\label{sec:random}

We have discussed the similarities and differences between bit and qubit as information carriers.  
As we have seen, in many ways they have the same properties, but in some more subtle ways qubit can be used to transmit information in a way that a bit is not capable of. 
The pressing question is if this difference is of any practical use.
In the following we describe one application.  

Alice and Bob want to generate sequences that disagree in every slot.  
Such sequences can be used by either Alice or Bob to authenticate messages to each other.  
For example, if Bob wants to send a message to Alice, he can attach part of his sequence to it.  
When Alice receives it, she can check that the subsequence differs from her corresponding subsequence in all slots.  
Of course, the same thing could be accomplished by sequences that agree in all of their slots.  
The disagreeing sequences become more interesting if there are more than two parties.  
Suppose that Alice has arranged for Bob and Charlie to have sequences that disagree with hers in every slot, but that Bob's and Charlie's sequences do not have this property.  
That allows Bob and Charlie to authenticate messages to Alice, but not to each other.  
Neither Bob nor Charlie can impersonate Alice to the other.
An example of such sequences are given in the table below.

\begin{center}
\begin{tabular}{|c!{\boldv}c|c|c|c|c|c|c|c|c|c|c|c|c|c|c|c|c|c|  } 
\boldh
Alice & b & c & b & c & a & c & a & b & a & c & b & c & c & b & c & a & c & c \\
\boldh
Bob  & c & b & c & a & b & a & c & c & c & b & a & a & b & a & b & c & a & b \\
\boldh
Charlie  & a & b & a & a & c & a & c & a & b & b & a & a & a & c & a & b & a & b \\
\boldh

\end{tabular}
\end{center}

How can Alice and Bob go about creating such sequences with no overlaps?  Suppose the symbols in the sequences are chosen from a set of four, which we shall call the alphabet $\{ a,b,c,d\}$.  Classically, the alphabet can be divided into two disjoint sets of two symbols each, $S_{0}=\{ a,b\}$ and $S_{1}=\{ c,d\}$.  For each element of the sequence, Alice picks a symbol from one set, and sends Bob a bit indicating from which set she chose.  Then Bob chooses an element from the other set.  This guarantees that their sequences will disagree. It does have a disadvantage, though.  Ideally, it would be best if all Alice knew about an element of Bob's sequence is that it is different from her corresponding element, and similarly for Bob.  For example, if her element were $a$, then all she would know about Bob's is that it is equally likely to be $b$, $c$, or $d$.  This gives each party the least information about the sequence of the other.  In the scheme described here that is not the case.  If, for example, Alice chooses $a$, then she tells Bob to choose an element in set $S_{1}$, that means she knows his element will be $c$ or $d$, but not $b$.  The situation becomes worse if the alphabet contains only 3 symbols, because then one of the sets only contains one element.  In that case, either Alice or Bob knows the other's element.

If Alice is sending a qubit instead of a bit, these problems disappear.  Consider a three symbol alphabet.  Alice can send a qubit to Bob in one of the three trine states
\begin{equation}
\begin{split}
|\psi_{a}\rangle &=  |0\rangle  \\
|\psi_{b}\rangle & =  -\frac{1}{2} |0\rangle + \frac{\sqrt{3}}{2} |1\rangle  \\
|\psi_{c}\rangle &  =  -\frac{1}{2} |0\rangle - \frac{\sqrt{3}}{2} |1\rangle .
\end{split}
\end{equation}
Bob measures the qubit using the measurement $\Mo$ defined as $\Mo(x) = \tfrac{2}{3}|\bar{\psi}_{x}\rangle\langle\bar{\psi}_{x}|$ ,  $x=a,b,c$, where
\begin{equation}
\begin{split}
|\bar{\psi}_{a}\rangle & =  |1\rangle  \\
|\bar{\psi}_{b}\rangle & =  -\frac{\sqrt{3}}{2} |0\rangle -\frac{1}{2} |1\rangle  \\
|\bar{\psi}_{c}\rangle & =  \frac{\sqrt{3}}{2} |0\rangle -\frac{1}{2} |1\rangle .
\end{split}
\end{equation}
Note that $\langle \psi_{x}|\psi_{y}\rangle = -1/2$ and $\langle\bar{\psi}_{x}|\bar{\psi}_{y}\rangle = -1/2$ for $x\neq y$, and $\langle \bar{\psi}_{x}|\psi_{x}\rangle = 0$.  Alice now sends a qubit to Bob in one of the trine states, and Bob measures the qubit using this measurement.  
Alice uses the label of the qubit state she sent to Bob as an element in her sequence and Bob uses the result of his measurement as an element of his.  Note that Bob has an equal chance of getting each of the symbols that Alice did not choose, hence this is an instance of uniform antidistinguishability.  
For example, if Alice chose $a$, then Bob can get either $b$ or $c$ with a probability of $1/2$.  Whichever one he gets, all he knows is that Alice's symbol is not the same as his.  Therefore, using qubits instead of bits allows us to achieve the goal of each party having the minimum amount of information possible about the sequence of the other.
We would like to note that a quantum-digital-signature protocol that makes use of measurements that eliminate possibilities, somewhat in the spirit of the measurements discussed here, has been proposed in \cite{wallden2015quantum}.

The quantum implementation of generating non-overlapping sequences has the feature that Alice does not know the sequences obtained by Bob and Charlie. We can, for instance, think of a situation where Alice is allowing them to vote on some occasion to be decided later. Alice wants to make sure that only Bob and Charlie can vote and no one else. 
Hence, she produces a random sequence for herself, encodes the symbols in the previously mentioned trine qubit states, and then sends them to Bob and Charlie. With the appropriate measurement, Bob and Charlie can produce random sequences that have the non-overlapping property with Alice's sequence, just like in the table above. With these sequences, they can sign their messages on a later occasion and, importantly for voting, they do not reveal their identity.

\section{Concluding remarks}\label{sec:conclusion}

We have been exploring communication scenarios in which Alice transmits a physical system to Bob, and they must have agreed on their preparation-measurement setup before the referee initiates the test. As we have seen, there are specific tasks where a qudit offers an advantage over a dit. 
Some remarks are in order. 

Firstly, it has been proven in \cite{FrWe15} that if Alice and Bob have access to a shared random number source, then they can implement with a dit system all communication matrices that they can implement with a qudit system.
The reason why shared randomness increases their capability is that it enables them to coordinate their choices of preparations and measurements from some preliminarily agreed sets, and in this way they can implement mixtures of those communication matrices that they can implement with a dit system. 
In mathematical terms, the sets of dit and qudit implementable communication matrices of the size $m\times n$ are different and non-convex \cite{heinosaari2020communication}, but their convex hulls are the same \cite{FrWe15}.
Practically, however, fully recovering the qudit communication matrices by dits and mixing would require an unlimited amount of shared randomness \cite{heinosaari2024simple}.
For that reason, we can conclude that the advantage of a qudit over a dit persists.

Secondly, in addition to asking about the minimal classical and quantum dimensions of specific communication matrices, we can ask for the most difficult communication tasks that can be achieved with a dit and qudit for some given $d$. 
There is an operationally motivated notion of difficulty, formulated as a relation called ultraweak matrix majorization \cite{HeKe19}.
In the set of communication matrices that can be implemented with a dit system, the most difficult one is the identity matrix $I_d$ that corresponds to the perfect communication of $d$ symbols. This single communication matrix can simulate all other dit implementable communication matrices.
As recently shown in \cite{heinosaari2023maximal}, the situation on the quantum side is much more complex and interesting. 
There is no qudit implementable communication matrix that is more difficult to implement than $I_d$.
However, there are infinitely many other communication matrices that do not compare with $I_d$, i.e., they are neither below nor above of $I_d$ in their difficulty. 
The practical usefulness of the corresponding communication tasks is yet to be found.

Thirdly, in the investigated setting we have been assuming that Bob's measurement device is fixed once Alice and Bob have decided their communication strategy. It is possible to allow Bob to vary his measurement device and in that way gather information in some unconventional way. 
This is exactly the case in quantum random access coding \cite{wiesner1983conjugate,ambainis2002dense}, which reveals a quantum advantage much in the same spirit as the uniform antidistinguishability does.
To achieve a benefit over a dit system one has to use incompatible quantum measurements \cite{CaHeTo20}, and in this way the advantage depends on collective quantum features of both preparations and measurements.
The broader communication scenario, where both Alice and Bob have control mechanisms, offers a wealth of research questions for exploration.

Overall, we conclude that a qudit can carry information differently than a dit. The full collection of practical applications that benefit from these diverse quantum characteristics remains to be discovered.

\end{document}